\documentclass[a4paper,11pt]{article}
\usepackage{pos}
\usepackage{graphicx}
\usepackage{cancel}

\graphicspath{{./figures/}}
\DeclareMathOperator*{\argmax}{argmax}

\title{A Monte Carlo estimator of flow fields for sampling and noise problems}

\author[a]{Michael S. Albergo}
\author*[b]{Gurtej Kanwar}

\affiliation[a]{Society of Fellows, Harvard University \\78 Mount Auburn St, Cambridge, MA United States}

\affiliation[b]{Higgs Centre for Theoretical Physics, School of Physics and Astronomy,\\
University of Edinburgh, EH9 3FD Edinburgh, United Kingdom}

\emailAdd{gkanwar@ed.ac.uk}

\abstract{Learned field transformations may help address ubiquitous critical slowing down and signal-to-noise problems in lattice field theory. In the context of an annealed sequence of distributions,
field transformations are defined by integrating flow fields that exactly solve a local transport problem. These proceedings discuss a new Monte Carlo approach to evaluating these flow fields, which can then be used directly in such contexts or as a means of generating unbiased training data for machine learning approaches.
By defining the Monte Carlo estimator using coupled Langevin noise, the statistical noise in the required integrals is significantly mitigated. Demonstrations of the method include a U(1) transport problem and an SU(N) glueball correlator.}

\FullConference{The 42nd International Symposium on Lattice Field Theory (LATTICE2025)\\
2-8 November 2025\\
Tata Institute of Fundamental Research, Mumbai, India\\}

\begin{document}
\maketitle

\section{Introduction}
We are interested in the problem of dynamically annealing from a simple prior density $\rho_0(\phi) := e^{-S_0(\phi) + F_0}$ through a sequence of densities parameterized by a fictitious time $t \geq 0$ as
\begin{equation} \label{eq:rho_t}
    \rho_t(\phi) := e^{-S_t(\phi) + F_t}, \qquad F_t = -\log\left( \int d[\phi] e^{-S_t(\phi)} \right).
\end{equation}
This can be achieved by deterministic evolution of the fields, starting from samples $\phi_{t=0}$ distributed according to $\rho_0(\phi)$, which are evolved in time via a \emph{flow},
\begin{equation}
\label{eq:ode}
    \dot{\phi}_t = b_t(\phi_t).
\end{equation}
The density of samples $\phi_t$ follow the distribution $\rho_t$, for all $t$, as long as the time-dependent vector field $b_t(\phi)$
satisfies the associated \emph{continuity equation},
\begin{equation} \label{eq:continuity}
    \nabla \cdot (b_t \rho_t) = -\partial_t \rho_t
    \quad 
    \implies \quad 
    \nabla \cdot b_t - b_t \cdot \nabla S_t = \partial_t S_t - \partial_t F_t.
\end{equation}

Depending on the choice of $\rho_t$, this paradigm has various use cases in the lattice field theory computational workflow. Most generally, the computation of expected values of operators under $\rho_t$ can be rewritten as an expectation value with respect to $\rho_0$ as
\begin{align} \label{eq:Ot}
\left< O \right>_t := \int d [\phi] \, O(\phi) \rho_t(\phi) =\int d [\phi_0] \, O(\phi_t) \rho_0(\phi_0),
\end{align}
where on the right side $\phi_t$ is defined by solving \eqref{eq:ode} with $\phi_0$ as the initial condition.
This change-of-variables relation has been explored as an option to reduce the cost of Monte Carlo integration~\cite{Luscher:2009eq,Engel:2011re,Bacchio:2022vje}, which may have significant computational benefits when one is interested in 
target distributions $\rho_t$ that are more challenging to simulate than the prior density $\rho_0$.

More recently, other works have explored applying a derivative with respect to $t$ in \eqref{eq:Ot} to insert operators into correlation functions~\cite{Bacchio:2023all,Abbott:2024kfc,Catumba:2025ljd}. Specifically, for correlation functions involving an operator $Q$, we define the time-dependent probability density in \eqref{eq:rho_t} using the action $S_t(\phi) = S(\phi) + t Q(\phi)$.
Differentiating any $n$-point correlation function with respect to this time dependence leads to a connected $(n+1)$-point correlation function,
\begin{equation}
\begin{aligned}
\partial_t \left< O_1 \dots O_n \right>_t \Big|_{t=0} = \left< O_1 \dots O_n Q \right>_0 - \left< O_1 \dots O_n \right>_0 \left< Q \right>_0.
\end{aligned}
\end{equation}
Assuming the flow field $b_t$ satisfies \eqref{eq:continuity}, the derivative at $t=0$ can be directly calculated as
\begin{equation} \label{eq:flow-obs}
    \partial_t \left< O_1 \dots O_n \right>_t \Big|_{t=0} 
    = \left< b_{t=0} \cdot \nabla (O_1 \dots O_n) \right>_0.
\end{equation}
This provides a separate approach to estimating the $(n+1)$-point correlation function, with the possibility of better signal-to-noise scaling.

In either the sampling or signal-to-noise context, if the flow field $b_t$ only approximately satisfies \eqref{eq:continuity}, exact corrective factors can be included to provide unbiased results. However, the computational benefits obtained in either case depend on the quality of the solution, and it is important to solve this high-dimensional partial differential equation to good accuracy. Several approaches have been devised to approximately solve for $b_t$ in the context of lattice field theory, including analytical small-$t$ expansion~\cite{Luscher:2009eq}, optimization of a physics-inspired ansatz~\cite{Bacchio:2022vje}, and machine learning parameterization and optimization~\cite{Bacchio:2023all,Albergo:2024trn}.

In this work, we propose a new Monte Carlo estimator for $b_t(\phi)$ which directly satisfies \eqref{eq:continuity} in expectation. While this introduces noise, an unbiased estimator of $b_t$ can be directly used in evaluating higher-point functions via the relation \eqref{eq:flow-obs}. It also serves as a ``ground truth'' estimate which can either be compared to machine learning outputs or used as training data for such methods. A complementary approach, based on stochastic automatic differentiation, was developed by Catumba and Ramos in Ref.~\cite{Catumba:2025ljd}; Section~\ref{sec:reln-stoch-ad} makes a concrete connection between the present work and the Catumba-Ramos method.

\section{Monte Carlo estimates of flow fields}

To begin, we write the flow field in terms of a time-dependent scalar potential $\varphi_t(\phi)$,
\begin{equation}
    b_t(\phi) = \nabla \varphi_t(\phi),
\end{equation}
in terms of which the continuity equation \eqref{eq:continuity} reduces to the nonlinear Poisson equation
\begin{equation} \label{eq:poisson}
    \nabla^2 \varphi_t - \nabla \varphi_t \cdot \nabla S_t = \partial_t S_t - \partial_t F_t.
\end{equation}
While still a challenging partial differential equation, a solution to \eqref{eq:poisson} is formally given for each value of $t$ by the Feynman-Kac formula~\cite{Kac1949:xxx},
\begin{equation} \label{eq:fk}
    \varphi_t(\phi) = - \int_0^\infty d\tau \, \mathbb{E}[\partial_t S_t(\Phi^\tau) - \partial_t F_t \,|\, \Phi^0 = \phi].
\end{equation}
Here, $\Phi^\tau$ is a random variable sampled by Langevin evolution over simulation time $\tau$ from the starting configuration $\Phi^0 = \phi$, according to\footnote{More precisely, this stochastic differential equation can be understood in the It\={o} sense as $d\Phi^\tau = - \nabla S_t d\tau + \sqrt{2} dW_\tau$ in terms of the Wiener process $W_\tau$.}
\begin{equation} \label{eq:fk-langevin}
    \partial_\tau \Phi^\tau = - \nabla S_t(\Phi^\tau) + \eta^\tau,
    \quad \left< \eta^\tau \eta^{\tau'} \right> = 2 \delta (\tau - \tau').
\end{equation}
The expectation value in \eqref{eq:fk} is taken over realizations of the stochastic noise $\eta^\tau$. A proof that this solves the relevant Poisson equation can for example be found in Ref.~\cite{Albergo:2024trn}.

\begin{figure}
    \centering
    \includegraphics{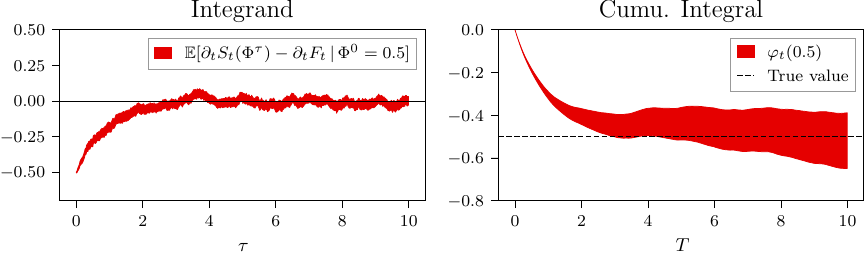}
    \caption{Evaluation of $\varphi_t(\phi=0.5)$ for the one-variable Gaussian distribution at $t = 0$ described in the main text. Results are shown for 1024 Langevin walkers $\Phi^\tau$ initialized from $\Phi^0 = 0.5$. The right plot compares the cumulative integral as a function of the maximum integration limit $T$ against the true value, demonstrating both convergence and the growing statistical noise expected from integrating the noisy integrand.}
    \label{fig:gaussian_fk_convergence}
    \vspace{0.5cm}

    \includegraphics{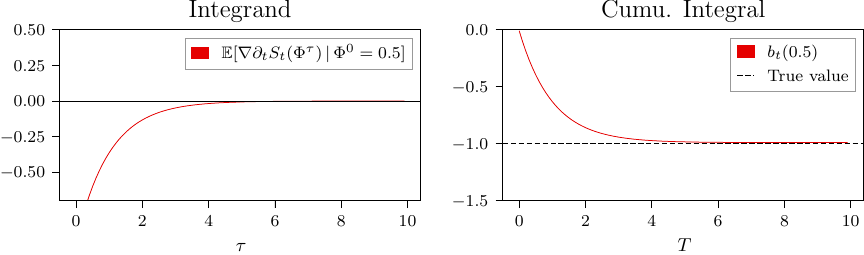}
    \caption{Evaluation of $b_t(\phi = 0.5) = \nabla \varphi_t(0.5)$ for the same theory as Figure~\ref{fig:gaussian_fk_convergence}. For this Gaussian theory, there is no statistical variation in $b_t$ using coupled Brownian noise. Evaluation is shown for one walker initialized from $\Phi^0 = 0.5$ and is compared against the true value.}
    \label{fig:gaussian_bt_convergence}
\end{figure}

Equation~\eqref{eq:fk-langevin} is the standard Langevin process for the action $S_t$~\cite{Batrouni:1985jn}, and the integrand therefore converges asymptotically to its expectation value under $S_t$. Under the assumption of geometric ergodicity, the integrand by definition converges exponentially to zero:
\begin{equation}
    \lim_{\tau \to \infty} \mathbb{E}[\partial_t S_t (\Phi^\tau) - \partial_t F_t] = \left< \partial_t S_t \right>_t - \partial_t F_t = 0.
\end{equation}
Equation \eqref{eq:fk} is then a well-defined integral representation for the potential and correspondingly the flow field. Practically speaking, however, integrating over noisy estimates of the expectation value leads to growing noise in this estimate of $\varphi_t$ as $\tau \to \infty$. This is illustrated in Figure~\ref{fig:gaussian_fk_convergence} for a simple time-dependent action $S_t = \tfrac{1}{2} \phi^2 + t \phi$ over a single variable $\phi \in \mathbb{R}$; in this example, the numerical results are compared against the exact solution $\varphi_t(\phi) = -\phi$ and $b_t(\phi) = -1$.

Despite growing noise in this estimator of \eqref{eq:fk}, we here propose to evaluate $b_t(\phi)$ by analytically differentiating \eqref{eq:fk} with respect to the initial condition $\Phi^0 = \phi$. This amounts to using linearity to move the gradient into the expectation value,
\begin{equation} \label{eq:coupled-bt}
\begin{aligned}
b_t(\phi) &= - \int_0^\infty d\tau \, \nabla_{\phi} \mathbb{E}[\partial_t S_t(\Phi^\tau) - \partial_t F_t \,|\, \Phi^0 = \phi] = - \lim_{T \to \infty} \mathbb{E}[\int_0^T \nabla_\phi \partial_t S_t(\Phi^\tau(\phi))].
\end{aligned}
\end{equation}
Above, $\Phi^\tau(\phi)$ is written as a function of the initial condition $\phi$, with respect to which the derivative is taken. For each Langevin noise sample $\eta^\tau$ to be used in the average, the gradient in \eqref{eq:coupled-bt} can for example be evaluated analytically by applying backpropagation to the numerical integration of $\hat{\varphi}_t^{[0,T]}$, where
\begin{equation}
    \hat{\varphi}^{[a,b]}:= -\int_a^b d\tau \, \partial_t S_t(\Phi^\tau).
\end{equation}
This gradient is straightforward to evaluate using standard automatic differentiation packages; Pytorch~\cite{Paszke:2019xhz} is used for all numerical results in this work.

The key insight is that differentiating with respect to the initial condition for fixed Langevin noise can yield low-noise estimates of $b_t(\phi)$ even when \eqref{eq:fk} leads to noisy estimates for $\varphi_t(\phi)$. This is demonstrated in Figure~\ref{fig:gaussian_bt_convergence} for the one-variable example, in which case this gives a \emph{zero-noise} estimator for $b_t(\phi)$. We note also the additional benefit of not having to estimate $\partial_t F_t = \left< \partial_t S_t \right>_t$ because it drops out under the gradient.

\begin{figure}
    \centering
    \includegraphics{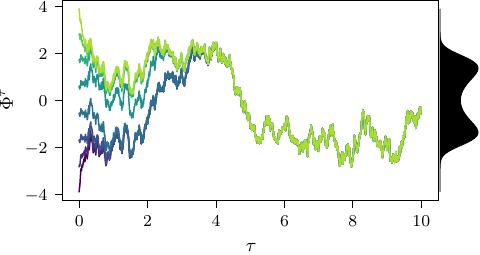}
    \hspace{0.5cm}
    \includegraphics{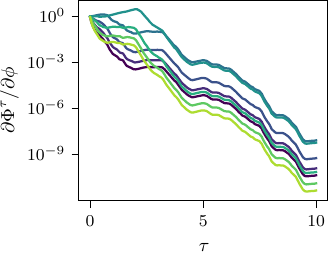}
    \caption{Left: Illustration of Langevin evolution with coupled noise in a quartic potential, showing eventual convergence of various initial conditions. Right: The dependence on the initial conditions, $\partial \Phi^\tau / \partial \phi$, which decreases exponentially when measured for the same trajectories.}
    \label{fig:coupled_demo}
\end{figure}

This estimate of the gradient conditioned on fixed samples of the noise process $\eta^\tau$ can be considered a specific case of the more general approach of \emph{coupled-noise} Langevin processes, in which correlated Langevin noise is used across different initial conditions. Remarkably, coupled-noise Langevin evolution for many physical potentials results in convergence to a path independent of the initial condition $\phi$ for \emph{each noise sample} $\eta^\tau$. An example of this behavior is shown for a simple quartic potential in Figure~\ref{fig:coupled_demo}, using a simple coupling where identical noise $\eta^\tau$ is applied for all initial conditions $\phi$. Using any coupled noise with this convergence property leads to the field dependence $\nabla_\phi \Phi^\tau$ vanishing asymptotically, as shown in the right panel of the figure. Evaluating \eqref{eq:coupled-bt} in this way translates to the integral converging asymptotically for each noise sample, suggesting an asymptotically constant growth of noise in estimates of $b_t(\phi)$ with this approach, if such a coupled Langevin process can be defined.

Finally, we describe an adjoint-sensitivity method of evaluating \eqref{eq:coupled-bt} without backpropagation~\cite{Li:2020sdg}. The \emph{adjoint state} at the reversed-time coordinate $\xi := (T-\tau)$ is defined to be
\begin{equation}
    A^\xi := \frac{\partial}{\partial \Phi^\xi} \hat\varphi_t^{[T-\xi,T]}. \end{equation}
It evolves according to a stochastic differential equation (SDE) for $\xi$ running from $0 \to T$,
\begin{equation} \label{eq:fk-sensitivity}
    \partial_{\xi} A^{\xi} = - \nabla \partial_t S_t(\Phi^{\xi}) - (A^\xi \cdot \nabla) \nabla \partial_t S_t(\Phi^\xi),
\end{equation}
where sample paths of $\Phi^\xi$ are obtained as the reverse of the sample paths $\Phi^\tau$.
The result in \eqref{eq:coupled-bt} is recovered by taking the expectation value and infinite-time limit,
\begin{equation}
    b_t(\phi) = \nabla \varphi_t = \lim_{T \to \infty} \mathbb{E}[A^{\xi=T}].
\end{equation}
While numerical results in the remainder of this work are obtained from the more direct backpropagation method, this adjoint sensitivity method is used for theoretical results below and may provide a more efficient route to evaluating $b_t$ in future work.

\section{Non-Euclidean domains}
We introduce two improvements that are helpful in the subsequent numerical demonstrations on compact, non-Euclidean domains.

First, as usual for Langevin evolution, \eqref{eq:fk-langevin} can be modified with a noise kernel $K$ to\footnote{This stochastic differential equation is to be understood in the It\={o} sense.}
\begin{equation} \label{eq:fk-langevin-kernel}
    \partial_\tau \Phi^\tau = -K(\Phi^\tau) \, \nabla S_t(\Phi^\tau) + \nabla \cdot K(\Phi^\tau) + \sqrt{K(\Phi^\tau)} \,\eta^\tau,
\end{equation}
without affecting the equilibrium distribution.
The solution $\tilde{b}_t(\phi)$ obtained by using  these modified dynamics in the evaluation of \eqref{eq:coupled-bt} satisfies a modified continuity equation
\begin{equation}
    \nabla \cdot (K \tilde{b}_t) - (K \tilde{b}_t) \cdot \nabla S_t = \partial_t S_t - \partial_t F_t.
\end{equation}
A valid flow field is then obtained in this case by assigning $b_t = K \tilde{b}_t$. Depending on the choice of kernel, the convergence and precision of estimates of $b_t$ may be improved. This approach is demonstrated below for a single $U(1)$ variable.

Second, we describe a specific coupling scheme for the Langevin noise in evaluating Feynman-Kac path integrals for $SU(N)$ theories. This approach is demonstrated below for alleviating the noise problem in a glueball correlator for a 2+1D $SU(N)$ lattice gauge theory.

\subsection{U(1) von Mises distribution}
\label{sec:u1-von-mises}

\begin{figure}
    \centering
    \includegraphics[width=0.8\textwidth]{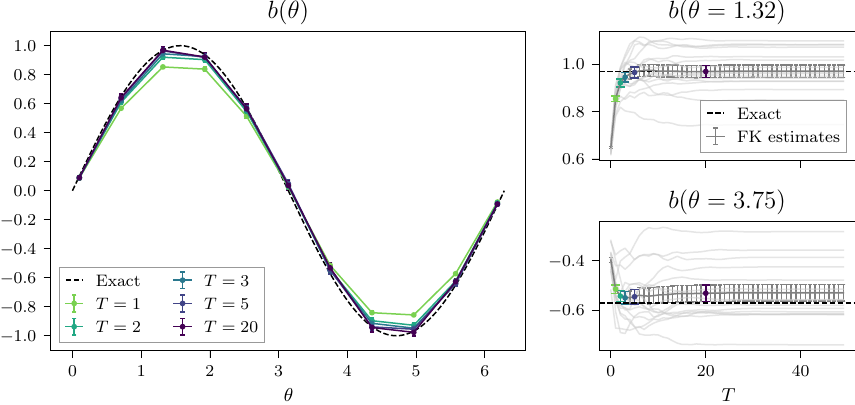}
    \caption{Estimate of $b(\theta)$ using the Feynman-Kac method for the von Mises distribution defined in the main text. Each point in the left panel is evaluated using 4096 Langevin walkers and is plotted for a range of upper integration limits $T$. The right panels show detailed convergence as a function of $T$ for two values of $\theta$. Evolution of block-averaged sample estimates are shown by light gray traces, and the bootstrap average over all walkers is indicated by the dark gray error bars.}
    \label{fig:u1_kernel_estimate}
\end{figure}

As a first example, we consider a single $U(1)$ variable $\theta \in [0,2\pi] $ with time-dependent action
\begin{equation}
    S_t(\theta) = t \cos(\theta),
\end{equation}
which is a von Mises distribution with concentration parameter $\kappa = -t$. At $t = 0$, the flow field can be analytically solved to give $b(\theta) := b_{t=0}(\theta) = \sin(\theta)$.
This exact solution is compared against a kernel-modified Feynman-Kac estimate in Figure~\ref{fig:u1_kernel_estimate}, with the noise kernel $K(\theta) = 0.5 + \sin(\theta/2)$,
which is chosen to be concave almost everywhere to improve the rate of convergence of the coupled-noise Langevin process.
As shown in the figure, the integral for $b_t$ converges \emph{per Monte Carlo sample}, giving asymptotically constant noise as the integration cutoff is removed. In contrast, we have found that evaluation without a kernel suffers from growing noise, requiring careful tuning of the cutoff time to balance systematic and statistical error.

\subsection{SU(N) glueball correlator}
As a second example, we apply the Feynman-Kac path integral to $SU(N)$ variables using coupled noise. 
In this context, the intrinsic positive curvature of the $SU(N)$ manifold provides an opportunity to design a useful coupling that is convergent even without a convex potential. Here, we choose the noise coupling to minimize the average distance between the evolving SU(N) variable $V^\tau$ and a reference variable $U^\tau$, where the Hilbert-Schmidt distance is used,
\begin{equation}
    d^2_{HS}(U^\tau, V^\tau) := 1 - \tfrac{1}{N}\mathrm{Re}\mathrm{Tr}[U^\tau (V^\tau)^\dagger].
\end{equation}
We first consider left-acting Brownian noise, which can be written in terms of Lie algebra components as $\eta^\tau = \eta^\tau_a T^a$. Any rotation $\xi^\tau_a = O_{ab} \eta^\tau_b$ preserves the correct statistics, so that a coupling can be designed by choosing the orthogonal matrix $O$ as a function of $U^\tau$ and $V^\tau$, with noise samples $\eta$ and $\xi$ respectively applied to $U^\tau$ and $V^\tau$. Considering the noise average of $d^2_{HS}(e^{i \eta^\tau \sqrt{d\tau}} U^{\tau}, e^{i \xi^\tau \sqrt{d\tau}} V^{\tau})$ for  $d\tau \to 0$, the rotation $O$ that minimizes this distance is~\cite{GowerProcrustes}
\begin{equation} \label{eq:ortho-coupling}
    O = \argmax_{O} \{ O^T M \} = M \sqrt{M^T M},
    \quad \text{where} \;\; M^{ab} := \mathrm{Re}\mathrm{Tr}[T^b T^a U^\tau (V^\tau)^\dagger].
\end{equation}
A right-acting, coupled Brownian noise can be defined by an analogous derivation.

The remaining choice is which trajectory or configuration $U^\tau$ to use as a reference. We here apply a \emph{self-coupling} strategy, since we are only interested in the derivative with respect to the initial condition: we take $U^\tau = \mathring{V}^\tau$ itself as the reference, where the circle indicates that we consider $\mathring{V}^\tau$ to be independent of the initial condition for the purpose of differentiating.\footnote{This can be implemented in Pytorch using \texttt{detach()}.}

\begin{figure}
    \centering
    \includegraphics[width=0.8\textwidth]{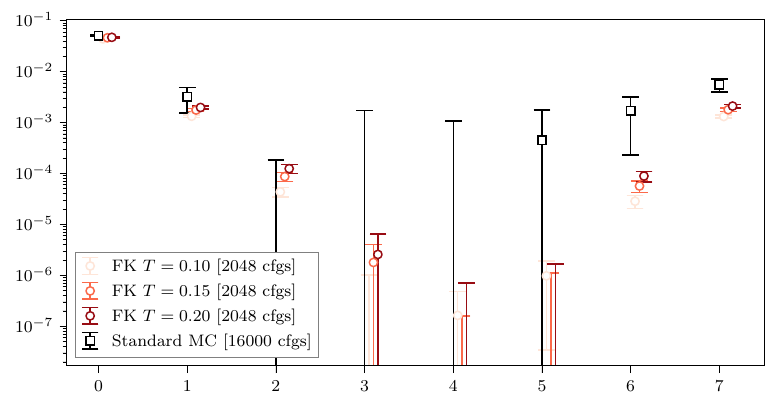}
    \caption{Evaluation of a $0^{++}$ glueball correlator as a function of separation $x/a$ for an $SU(2)$ lattice gauge theory as described in the main text. Estimates using a standard vacuum-subtracted estimator are compared to the flow-field approach with a Feynman-Kac estimator integrated up to a variety of integration limits $T$.}
    \label{fig:sun_glueball}
\end{figure}

To test this approach, we evaluate a $0^{++}$ glueball correlator in an $SU(2)$ Yang-Mills theory with Wilson gauge action at inverse gauge coupling $\beta = 4.0$ on a $4^2 \times 8$ lattice. To improve convergence of the Feynman-Kac integral for $b_t$ in this context, we apply the $SU(N)$ coupled noise defined above acting on both the left and right side of each gauge link. We further apply a constant kernel, $K = \alpha P + (1-P)$, to rescale noise along the pure-gauge directions by a factor $\alpha = 10$, where $P$ is a projector onto the pure-gauge direction acting in the space of left and right acting noise at each site $x$.
The observable is determined by applying the derivative trick to a one-point function,
\begin{equation}
    \left< O(x) O(0) \right> - \left< O(x) \right> \left< O(0) \right> = \left< b_{t=0} \cdot (\nabla O(x) O(0)) \right>,
\end{equation}
where the glueball interpolator is chosen to be a local plaquette operator $O(x) = \mathrm{Re}\mathrm{Tr}[P_{01}(x)]$.
The results are compared to a standard vacuum-subtracted estimate in Figure~\ref{fig:sun_glueball}; significantly more precise results are obtained, while using approximately $8\times$ fewer configurations.

\section{Relation to Catumba-Ramos}
\label{sec:reln-stoch-ad}
A similar method to estimate flow fields was recently introduced by Catumba and Ramos in Ref.~\cite{Catumba:2025ljd}, in which the final state of a continuous Markovian process is differentiated with respect to the time dependence of the action. This can be related to the present work by taking the Markovian process in their method to be an overdamped Langevin simulation for the action $S_t$ with a fixed choice of parameter $t$,
\begin{equation} \label{eq:fwd-langevin}
    \partial_\xi \bar{\Phi}^\xi = -\nabla S_t(\bar{\Phi}^\xi) + \eta^\xi.
\end{equation}
Here we use the barred variable $\bar{\Phi}^\xi$ to distinguish from the Langevin evolution $\Phi^\tau$ used above and suggestively use $\xi$ for Langevin time in this context. The sensitivity of $\bar\Phi^\xi$ to the $t$-dependence of the action follows an SDE,
\begin{equation} \label{eq:cr-sensitivity}
    \bar{A}^\xi := \tfrac{d}{dt} \bar\Phi^\xi
    \quad \implies \quad
    \partial_\xi \bar{A}^\xi = - \nabla \partial_t S_t(\bar\Phi^\xi) - (\bar{A}^\xi \cdot \nabla) \nabla S_t(\bar\Phi^\xi).
\end{equation}
For a fixed starting distribution for $\bar\Phi^0$, ergodicity implies that the Markov chain evolution of $\bar\Phi^\xi$ will converge to the distribution $\rho_t$ for $\xi \to \infty$. Perturbing the action appearing in the Langevin evolution to $S_{t+dt}$ will similarly result in convergence to the distribution $\rho_{t+dt}$, such that a solution for the flow field that corresponds to this change in density is given by the conditional expectation value
\begin{equation} \label{eq:bt-stoch-ad}
    b_t(\phi) = \lim_{T \to \infty} \mathbb{E}[\tfrac{d}{dt} \bar\Phi^T \,|\, \bar\Phi^T = \phi] = \lim_{T \to \infty} \mathbb{E}[\bar{A}^{\xi=T}].
\end{equation}

It is clear from \eqref{eq:fk-sensitivity} and \eqref{eq:cr-sensitivity} that there is an intimate connection between the Catumba-Ramos approach and the Feynman-Kac determination of the flow field.
Here we sketch a proof that these formulations are in fact equivalent under reversal of Langevin time $\tau \leftrightarrow \xi$, when the starting condition for the former is the equilibrium distribution, $\rho_t$. The equivalence of these formulations is immediate if the following two distributions are equal to each other
\begin{enumerate}
    \item The distribution of paths $\Phi^\xi$ obtained by reversing the paths $\Phi^\tau$ generated by the standard Langevin simulation in \eqref{eq:fk-langevin}, starting from $\Phi^{0} = \phi$ and evolving to time $\tau = T$.
    \item The conditional distribution of paths $\bar\Phi^\xi$ generated by the (same) Langevin simulation in \eqref{eq:fwd-langevin}, starting from $\bar\Phi^0$ sampled from the equilibrium distribution $\rho_t$ and conditioning on the final state $\bar\Phi^{T} = \phi$.
\end{enumerate}
We show equivalence of these distributions using detailed balance, $p(A \to B) = \tfrac{\rho_t(B)}{\rho_t(A)} p(B \to A)$, where $p(\cdot)$ is the transition kernel for Langevin evolution with timestep $dt = T/N$, and we use the fact that $\rho_t$ is the equilibrium distribution.\footnote{Though typical discretizations of Langevin evolution do not exactly satisfy detailed balance, in the limit $N \to \infty$ corrections from violating detailed balance go to zero.}
The pathwise probabilities forward and backwards can then be equated in the continuous-time limit,
\begin{equation}
\begin{aligned}
    p(\bar\Phi^\xi) &= \lim_{N \to \infty} p(\bar\Phi^0 \to \bar\Phi^{dt}) p(\bar\Phi^{dt} \to \bar\Phi^{2 dt}) \dots \frac{\rho_t(\bar\Phi^0) \delta(\bar\Phi^T - \phi)}{\rho_t(\phi)} \\
    p(\Phi^\xi) &= \lim_{N \to \infty} p(\Phi^{dt} \to \Phi^0) p(\Phi^{2 dt} \to \Phi^{dt}) \dots \delta(\Phi^T - \phi) \\
    &= \lim_{N \to \infty} \tfrac{\rho_t(\Phi^0)}{\cancel{\rho_t(\Phi^{dt})}} p(\Phi^0 \to \Phi^{dt}) \tfrac{\cancel{\rho_t(\Phi^{dt})}}{\cancel{\rho_t(\Phi^{2 dt})}} p(\Phi^{dt} \to \Phi^{2 dt}) \dots \delta(\Phi^T - \phi) = p(\bar\Phi^\xi).
\end{aligned}
\end{equation}

Under time reversal of the sampled processes, our approach is thus statistically equivalent to the method of Ref.~\cite{Catumba:2025ljd}, when overdamped Langevin is used. We however highlight that working with our formulation allows directly estimate the flow field for fixed input field configurations, as shown for example in Sec.~\ref{sec:u1-von-mises}. Moreover, framing the convergence problem in terms of convergence of coupled Langevin trajectories provides a new perspective that may yield progress on the challenges of evaluating flow fields in gauge theories.

\section{Conclusions}
This work introduced a Monte Carlo estimator for the flow field $b_t$ that transports samples under a time-dependent action $S_t$. Such flow fields may be useful for both ensemble generation and noise problems in observables. By incorporating specific kernel transformations and an $SU(N)$ coupled-noise process, our method of estimating $b_t$ was shown here to be effective for a $U(1)$ von Mises distribution and for measurements of a glueball correlation function in 2+1D $SU(N)$ Yang-Mills.

We have also demonstrated a close relation between our approach and the stochastic automatic differentiation method of Ref.~\cite{Catumba:2025ljd}. While the overdamped Langevin dynamics used here generally have slower mixing time than the Hamiltonian evolution which has been recently applied in that context, the deep connection between the two approaches also suggests a Hamiltonian version of the Feynman-Kac path integral, which will be investigated in future work.

In the future, these Monte Carlo estimates of flow fields may provide useful training data for complementary machine learning or analytical approaches seeking to parameterize and learn the same flow fields. High-quality deterministic parameterizations based on this approach would further improve on the sampling and noise-reduction benefits conferred by using flow transport.

\acknowledgments
We are grateful to Ryan Abbott and Alberto Ramos for insightful discussions.  MSA is supported by a Junior Fellowship at the Harvard Society of Fellows as well as the National Science Foundation under Cooperative Agreement PHY-2019786 (The NSF AI Institute for Artificial Intelligence and Fundamental Interactions\footnote{http://iaifi.org/}). This work has been made possible in part by a gift from the Chan Zuckerberg Initiative Foundation to establish the Kempner Institute for the Study of Natural and Artificial Intelligence, and was supported in part by the Schmidt Sciences AI2050 Early Career Fellowship.

\bibliographystyle{abbrv}
\bibliography{main}

\end{document}